\documentclass{iopart}
\pdfoutput=1
\usepackage{euscript,iopams,amssymb,amsfonts,graphicx,bm}
\usepackage{setstack}

  \usepackage{epstopdf}
  \usepackage{graphics}

\usepackage{color}

\eqnobysec

\newcommand{\p}{\mathbf{p}}

\newcommand{\Markov}[2]{\underset{#1}{\overset{#2}{\rightleftharpoons}}}
\newcommand{\x}{\mathbf{x}}

\renewcommand{\e}{{\rm e}}

\def\P{{\mathbb P}} 
 
\def\R{{\mathbb R}}

\renewcommand{\theequation}{\arabic{section}.\arabic{equation}}

\begin{document}

\title{Spin coherent states and stochastic hybrid path integrals} 
\author{Paul C. Bressloff$^{1}$ }
\address{$^1$Department of Mathematics, University of Utah, 155 South 1400 East, Salt Lake City, Utah 84112, USA}
\ead{bressloff@math.utah.edu}

\date{\today}

\begin{abstract} Stochastic hybrid systems involve a coupling between a discrete Markov chain and a continuous stochastic process. If the latter evolves deterministically between jumps in the discrete state, then the system reduces to a piecewise deterministic Markov process (PDMP). Well known examples include stochastic gene expression, voltage fluctuations in neurons, and motor-driven intracellular transport. In this paper we use coherent spin states to construct a new path integral representation of the probability density functional for stochastic hybrid systems, which holds outside the weak noise regime. We use the path integral to derive a system of Langevin equations in the semi-classical limit, which extends previous diffusion approximations based on a quasi-steady-state reduction. We then show how in the weak noise limit the path integral is equivalent to an alternative representation that was previously derived using Doi-Peliti operators. The action functional of the latter is related to a large deviation principle for stochastic hybrid systems.

 \end{abstract}


\maketitle

\section{Introduction}\label{sec:Intro}

Stochastic hybrid systems, which involve a coupling between a discrete Markov chain and a continuous stochastic process, are finding an increasing number of applications in biology \cite{Bressloff14a,Bressloff17a}. One of the simplest examples is a velocity jump process describing a particle randomly switching between different velocity states. The particle could represent a bacterial cell undergoing chemotaxis \cite{Berg77,Hillen00} or a motor-cargo complex walking along a cytoskeletal filament \cite{Reed90,Friedman05,Newby10,Newby10a,Bressloff13a}. A second example is a conductance-based model of a neuron \cite{Fox94,Chow96,Keener11,Goldwyn11,Buckwar11,NBK13,Bressloff14b,Newby14}, which considers the dynamics of the membrane voltage under the stochastic opening and closing of membrane-bound ion channels whose transition rates are voltage dependent. The number of open ion channels at time $t$ is represented by a discrete random variable $N(t)$ and the voltage by a continuous random variable $X(t)$. Another important application is stochastic gene expression, where $N(t)$ could represent the activity state of a gene (due to the binding/unbinding of transcription factors) and $X(t)$ the concentration of a synthesized protein \cite{Kepler01,Bose04,Newby12,Newby15,Hufton16}. (Note, however, that in the case of low protein concentrations, one has to keep track of the protein number, which is modeled as a second discrete process \cite{Sasai03,Zhang13,Bhatt20}.) A final example is a stochastic hybrid neural network of synaptically coupled neuronal populations \cite{Bressloff13a,Bressloff15,Yang19}; the state of each local population is described in terms of two stochastic variables, a continuous synaptic variable and a discrete activity variable.

A piecewise deterministic Markov process (PDMP) is a particular subset of stochastic hybrid systems, in which the continuous random variables evolve deterministically between jumps in the discrete random variables \cite{Davis84}. PDMPs have been studied extensively within the context of large deviation theory \cite{Kifer09,fagg09,Faggionato10,Bressloff17}. One major finding of these studies is that the rate function  of the associated large deviation principle (LDP) can be related to an action functional, whose Hamiltonian corresponds to the principal eigenvalue of a linear operator. The latter incorporates both the generator of the discrete Markov process and the vector fields of the piecewise deterministic dynamics. An alternative method for deriving the action is to construct a path integral representation of the probability density functional in the weak noise limit. We originally derived such a path integral using integral representations of Dirac delta functions \cite{Bressloff13a,Bressloff15}, analogous to the analysis of stochastic differential equations (SDEs) \cite{Martin73,Dom76,Janssen76}. (This approach, which avoids the use of ``quantum-mechanical'' operators, can also be applied to master equations by considering differential equations for the corresponding generator or marginalized distribution of the Markov process \cite{Weber17}.) Recently, we developed a more efficient and flexible framework for constructing hybrid path integrals in the weak noise limit \cite{Bressloff21}, which combines the Doi-Peliti operator formalism for master equations \cite{Doi76,Doi76a,Peliti85,Weber17} with an analogous operator method for SDEs \cite{Holmes20}.

One of the major steps in the derivation of the Doi-Peliti path integral for master equations is to project the discrete states onto an overdetermined set of coherent ``bosonic'' states. This is particularly useful when the number of discrete states is unbounded, as in a variety of birth-death processes. However, when the number of discrete states is two or three, say, then a more natural decomposition is in terms of coherent ``spin'' states \cite{Radcliffe71,Fradkin13}. Such a decomposition has recently been used to study stochastic gene expression in the presence of promoter noise and low protein copy numbers \cite{Sasai03,Zhang13,Bhatt20}. For example, one can effectively map the stochastic dynamics of a single genetic switch to a quantum spin-boson system. In this paper we use coherent spin states to construct a hybrid path integral for stochastic hybrid systems with a small number of discrete states, which holds outside the weak noise limit. We then use the path integral to derive a system of Langevin equations in the semi-classical limit, which extends previous diffusion approximations based on a quasi-steady-state reduction. We also show how in the weak noise limit the path integral reduces to the alternative representation that was previously derived using Doi-Peliti operators.

The structure of the paper is as follows. In section 2 we formulate a two-state stochastic hybrid system whose probability density evolves according to a differential Chapman-Kolmogorov (CK) equation. We also present a few applications. The operator formalism involving coherent spin states is introduced in section 3, which is used to derive an operator version of the CK equation for the two-state model. We also indicate how to extend the theory to a three-state model. In section 4 we derive the hybrid path integral and consider the semi-classical limit. The alternative path integral representation is summarized in section 5, and the relationship between the two representations is established in section 6.

\section{Two-state stochastic hybrid systems}

In order to develop the basic theory, consider a stochastic hybrid system whose state at time $t$
consists of the pair $(X(t),N(t))$, where $X(t)\in \R$ and $N(t)\in \{0,1\}$. Suppose that the discrete process evolves according to the two-state Markov chain
\begin{equation}
0\Markov{\alpha} {\beta}1.
\end{equation} 
We also allow the transition rates to depend on the continuous state variable $X(t)$, that is,
$\alpha=\alpha(x),\beta=\beta(x)$ for $X(t)=x$. In between jumps in the discrete variable, $X(t)$ evolves according to the Ito SDE
\begin{equation}
\label{pd}
dX=F_n(x)dt+\sqrt{2D_n(x)}dW
\end{equation}
for $N(t)=n$, where $W(t)$ is a Wiener process with
\[\langle W(t)\rangle =0,\quad \langle W(t)W(t')\rangle = \min\{t,t'\}.\]
Introduce the probability density 
\begin{eqnarray}
  &P_n(x,t)dx=\mbox{Prob}\{X(t)\in (x,x+dx), N(t)=n\},
\end{eqnarray}
given an initial state $X(0)=x_0,N(0)=n_0$.
The probability density evolves according to the differential CK equation
\begin{eqnarray}
\label{CK}
\fl \frac{\partial P_n(x,t)}{\partial t}=-\frac{\partial F_n(x)P_n(x,t)}{\partial x}+\frac{\partial^2 D_n(x)P_n(x,t)}{\partial x^2}+\sum_{m= 0,1}Q_{nm}(x)P_m(x,t),
\end{eqnarray}
with matrix generator
\begin{equation}
\label{Q}
{\bf Q}=\left (\begin{array}{cc} -\beta(x) &\alpha(x) \\ \beta(x)& -\alpha(x) \end{array} \right ).
\end{equation}

\begin{figure}[t!]
\raggedleft
\includegraphics[width=8cm]{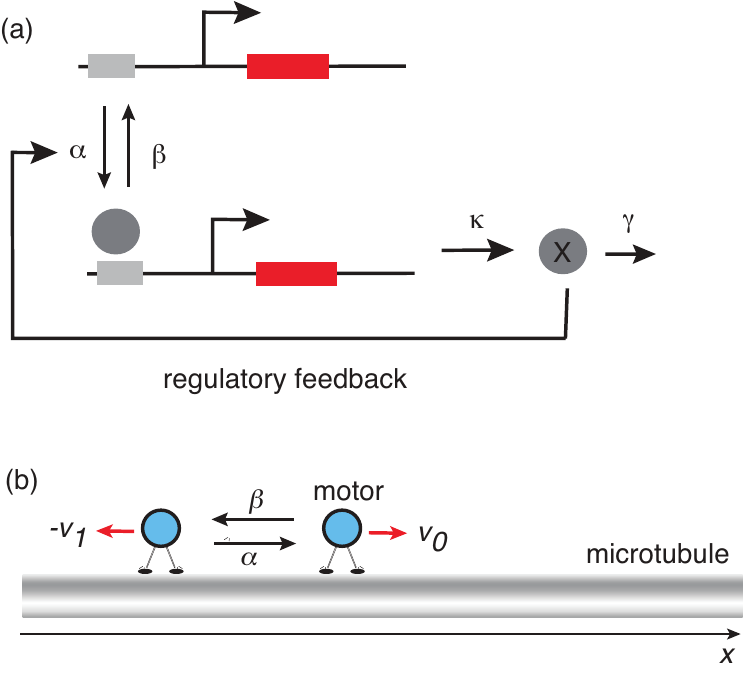}
\caption{Examples of two-state hybrid systems. (a) An autoregulatory network. A gene is activated (or repressed) by its own protein product $X$ when it binds to a promoter site. (b) Bidirectional motor transport.}
\label{auto}
\end{figure}

One of the simplest examples of a two-state hybrid system is a gene network with autoregulatory feedback, see Fig. 
\ref{auto}(a). Let $x(t)$ denote the concentration of protein $X$ at time $t$ and let $N(t)$ represent the current state of the gene. If $N(t)=0$ then the gene is active and synthesizes the protein at a rate $\kappa$, whereas if $N(t)=1$ then the gene is inactive and protein production halts.  We thus have the PDMP
\begin{equation}
\frac{dx}{dt}=F_n(x)\equiv  \kappa (1-n) -\gamma x,
\label{autoreg}
\end{equation}
where $\gamma$ is the protein degradation rate. Now suppose that the gene is active when one of its operator sites is bound by $X$ and inactive when it is unbound. Switching between the inactive and active states is then controlled by protein binding/unbinding. In particular, we can identify $\alpha$ and $\beta$ with the binding and unbinding rates, respectively. Moreover, $\alpha$ will depend on the protein concentration $x$ due to the autoregulatory feedback. The kinetic equation (\ref{autoreg}) ignores any fluctuations in the number of proteins due to finite-size effects. Let the number of proteins at time $t$ be $M(t)=x(t)\Omega$ where $\Omega$ is the system-size (cell volume, say). If we include both promoter and protein fluctuations, then the stochastic dynamics is described by a master equation for the joint probability distribution $P(n,m,t)=\P[N(t)=n,M(t)=m]$, 
where $m\geq 0$ and $n\in \{0,1\}$ \cite{Sasai03,Zhang13,Hufton16}:
\begin{eqnarray}
\label{greg0}
\fl&\frac{dP(n,m,t)}{dt}=[\beta n+\alpha (1-n)]P(1-n,m,t)+\kappa (1-n)  P(n,m-1,t)\\
\fl&\quad +\gamma (m+1)P(n,m+1,t)-\left [\beta(1-n)+\alpha n+\kappa(1-n) +\gamma  m\right ]P(n ,m,t),\nonumber
\end{eqnarray}
with $P(n,-1,t)=0$. A stochastic hybrid system of the form (\ref{pd}) can then be obtained by carrying out a system-size expansion of the master equation with
\begin{equation}
F_n(x)=\kappa (1-n)-\gamma x,\quad D_n(x)=\Omega^{-1}\left (\kappa (1-n)+\gamma x\right ).
\end{equation}

A second example of a two-state hybrid system is bidirectional motor transport within cells \cite{Newby10,Newby10a,Bressloff13}. Consider a particle moving along a one-dimensional track, see Fig. \ref{auto}(b). The particle could represent a motor-cargo complex and the track could represent a set of microtubular filaments in the axon of a neuron.\footnote{Microtubules are polarized polymeric filaments with biophysically distinct ($+$) and $(-)$ ends, and this polarity determines the preferred direction in which an individual molecular motor moves. For example, kinesin moves towards the $(+)$ end whereas dynein moves towards the $(-)$ end. One mechanism for bidirectional transport is a tug-of-war model between opposing groups of processive motors \cite{Gross04,Muller08}.} The particle randomly switches between a right-moving (anterograde) state with speed $v_0$ and a left-moving (retrograde) state with speed $v_1$. If $N(t)$ denotes the velocity state at time $t$, then the position $x(t)$ of the motor evolves according to the simple PDMP
\begin{equation}
\label{ratem}
\frac{dx}{dt}=(1-2n) v_n
\end{equation}
for $N(t)=n$. One mechanism for generating an $x$-dependent transition rate involves microtubule associated proteins (MAPs). These molecules bind to microtubules and effectively modify the free energy landscape of motor-microtubule interactions.  For example, tau is a MAP found in the axon of neurons and is known to be a key player in Alzheimer's disease. Experiments have shown that tau can significantly alter the dynamics of kinesin; specifically, by reducing the rate at which kinesin binds to the microtubule \cite{Vershinin07}. Thus tau signaling can be incorporated into a motor transport model by considering a tau concentration-dependent kinesin 
binding rate \cite{Newby10a}. In terms of the simplified two-state PDMP, this means that the rate of switching to the right-moving state becomes a decreasing function of the $\tau$ concentration $c$. Assuming that the latter varies with $x$, we have $\alpha(x)=\alpha(c(x)), \alpha'(c)<0$. Finally, one obtains a stochastic hybrid system of the form (\ref{pd}) if one also takes into account a diffusive component in the motion of the particle.

\section{Operator formalism}

Recently we combined operator formulations of master equations \cite{Doi76,Doi76a,Peliti85} and SDEs \cite{Holmes20} to rewrite the CK equation (\ref{CK}) as an operator equation acting on a Hilbert space \cite{Bressloff21}. This was then used to derive a corresponding hybrid path integral in the weak noise limit. In order to derive a path integral that holds for arbitrary levels of noise, we consider an alternative operator formalism that is particularly useful when the number of discrete states is small. The basic idea is to replace the bosonic annihilation and creation operators of Doi-Peliti with Pauli spin operators acting on coherent spin states. The latter have also recently been used to study the effects of promoter noise in gene networks \cite{Sasai03,Zhang13,Bhatt20}. For the sake of clarity, we introduce the continuous and discrete operator constructions separately, and then show to combine them in the case of the hybrid system. (For ease of notation, we suppress the $x$-dependence of the transition rates However, all of the results hold if such $x$-dependence is included, unless stated otherwise. We also take $x\in \R$ although the actual dynamics may restrict the domain of $x$. For example, in the  case of a gene network $x$ represents a concentration so that we require $x\geq 0$.)

\subsection{Fokker-Planck equation} Consider the following Ito SDE for $X(t)\in \R$:
\begin{equation}
dX(t)=A(X)dt+\sqrt{2D(X)}dW(t).
\end{equation}
The corresponding Fokker-Planck (FP) equation for the probability density $P(x,t)$ is
\begin{equation}
\frac{\partial P(x,t)}{\partial t}=-\frac{\partial A(x)P(x,t)}{\partial x}+\frac{\partial^2 D(x)P(x,t)}{\partial x^2}.
\end{equation}
\numparts
Following Ref. \cite{Holmes20,Bressloff21}, we introduce a Hilbert space spanned by the vectors $|x\rangle$, together with a conjugate pair of position-momentum operators $\hat{x}$ and $\hat{p}$ such that
\begin{equation}
\label{com2}
[\hat{x},\hat{p}]=i.
\end{equation}
Their action on the given Hilbert space is taken to be
\begin{equation}
\hat{x}|x\rangle = x|x\rangle,\quad \hat{p}|x\rangle=-i\overset{\leftarrow}{\frac{d}{d x}}|x\rangle.
\end{equation}
The arrow on the differential operator indicates that it operates to the left. Alternatively, given a state vector $|\phi\rangle =\int_{-\infty}^{\infty}dx\phi(x) |x\rangle$, we have $\langle x|\hat{p}|\phi\rangle =-i\phi'(x)$.
 \endnumparts
The inner product and completion relations on the Hilbert space are
\begin{equation}
\label{fpcom1}
\langle x'|x\rangle =\delta(x-x'),\quad  \int_{-\infty}^{\infty} dx\, |x\rangle \langle x|=1.
\end{equation}
Given the probability density $P(x,t)$ we define the state vector 
\begin{equation}
|\psi(t)\rangle = \int_{-\infty}^{\infty} dx\, P(x,t)|x\rangle .
\end{equation}
Differentiating both sides with respect to time $t$ and using the FP equation gives
\begin{eqnarray*}
\frac{d}{dt}|\psi(t)\rangle&= \int_{-\infty}^{\infty} dx\,\left [-\frac{\partial A(x)P(x,t)}{\partial x}+\frac{\partial^2 D(x)P(x,t)}{\partial x^2}\right ]|x\rangle\\
&= \int_{-\infty}^{\infty} dx\,\left [-A(x)P(x,t)\overset{\leftarrow}{\frac{\partial }{\partial x}}+ D(x)P(x,t)\overset{\leftarrow}{\frac{\partial^2}{\partial x^2}}\right ]|x\rangle\\
&=\left [-i\hat{p}A(\hat{x})-\hat{p}^2 D(\hat{x})\right ]\int_{-\infty}^{\infty} dx\, P(x,t)|x\rangle.
\end{eqnarray*}
Hence, we can write the FP equation in the operator form
\numparts
\begin{equation}
\frac{d}{dt}|\psi(t)\rangle=\hat{H}_{\rm fp} |\psi(t)\rangle,
\end{equation}
with
\begin{equation}
\label{fp:H}
\hat{H}_{\rm fp}=-i\hat{p}A(\hat{x})-\hat{p}^2 D(\hat{x}).
\end{equation}
\endnumparts
The formal solution of the FP equation is
\begin{equation}
|\psi(t)\rangle=\e^{\hat{H}_{\rm fp}t}|\psi(0)\rangle,
\end{equation}
and expectations are given by
\begin{equation}
\fl \langle X(t)\rangle = \int_{\infty}^{\infty}dx\, xP(x,t)=\int_{-\infty}^{\infty} dx\, \langle x|\hat{x}|\psi(t)\rangle=\int_{-\infty}^{\infty} dx\, \langle x|\hat{x}\e^{\hat{H}_{\rm fp}t}|\psi(0)\rangle.
\end{equation}

Another useful choice of basis vectors is the momentum representation (analogous to taking Fourier transforms),
\begin{equation}
|p\rangle =\int_{-\infty}^{\infty}dx\,  \e^{ipx}|x\rangle.
\end{equation}
It immediately follows that $|p\rangle$ is an eigenvector of the momentum operator $\hat{p}$, since
\begin{equation}
\hat{p}|p\rangle=\int_{-\infty}^{\infty}dx\,  \e^{ipx}\left (-i\overset{\leftarrow}{\frac{d}{d x}}\right )|x\rangle=\int_{-\infty}^{\infty}dx\, p \e^{ipx}|x\rangle =p|p\rangle.
\end{equation}
Using the inverse Fourier transform, we also have
\begin{equation}
|x\rangle =\int_{-\infty}^{\infty}\frac{dp}{2\pi}\,  \e^{-ipx}|p\rangle,
\end{equation}
and the completeness relation
\begin{equation}
\label{fpcom2}
\int_{-\infty}^{\infty}\frac{dp}{2\pi}|p\rangle \langle p|=1.
\end{equation}

\subsection{Coherent spin states}

Consider the master equation for a two-state Markov chain, written in matrix form
\begin{equation}
\label{mc}
\frac{d{\bf P}}{dt}={\bf Q}{\bf P}(t),\quad {\bf P}(t)=(P_0(t),P_1(t))^{\top},
\end{equation}
with ${\bf Q}$ the matrix (\ref{Q}). Introduce the Pauli spin matrices
\begin{equation}
\fl \sigma_x=\frac{1}{2}\left (\begin{array}{cc} 0&1 \\ 1& 0 \end{array} \right ),\quad \sigma_y=\frac{1}{2}\left (\begin{array}{cc} 0&-i \\ i& 0 \end{array} \right ),
\quad \sigma_z=\frac{1}{2}\left (\begin{array}{cc} 1&0 \\ 0& -1 \end{array} \right ),
\end{equation}
and set
\begin{equation}
\sigma_{\pm}=\sigma_x\pm i\sigma_y.
\end{equation}
It follows that the generator can be rewritten as
\begin{equation}
{\bf Q} = -\beta \left (\frac{1}{2}{\mathbf 1} +\sigma_z\right) -\alpha \left (\frac{1}{2}{\mathbf 1} -\sigma_z\right) +\alpha \sigma_++\beta\sigma_-.
\end{equation}
Next we define the coherent spin-$1/2$ state \cite{Radcliffe71,Sasai03,Zhang13,Bhatt20}
\begin{equation}
|s\rangle = \left (\begin{array}{c} \e^{i\phi/2}\cos^2\theta/2  \\ \e^{-i\phi/2}\sin^2\theta/2 \end{array} \right ),\quad 0\leq \theta \leq \pi,\ 0\leq \phi <2\pi,
\end{equation}
together with the adjoint
\begin{equation}
\langle s|=\left ( \e^{-i\phi/2}  ,\,  \e^{i\phi/2}  \right ).
\end{equation}
Note that 
\begin{equation}
\langle s'|s\rangle =\e^{i(\phi-\phi')/2}\cos^2\theta/2+\e^{-i(\phi-\phi')/2}\sin^2\theta/2,
\end{equation}
so that $\langle s|s\rangle =1$ and
\begin{equation}
\label{sdiff}
\langle s+\Delta s |s\rangle =1-\frac{1}{2}i \Delta \phi \cos \theta +O(\Delta \phi^2).
\end{equation}
We also have the completeness relation
\begin{equation}
\label{comp2}
\frac{1}{2\pi} \int_0^{\pi}\sin \theta \, d\theta \int_0^{2\pi}d\phi\, |s\rangle \langle s|=1.
\end{equation}
It can checked that the following identities hold:
\numparts
\begin{eqnarray}
\langle s|\sigma_z|s\rangle &=\frac{1}{2}\cos \theta, \\
\langle s|\sigma_+|s\rangle &= \frac{1}{2}\e^{i\phi}\sin\theta, \\
\langle s|\sigma_-|s\rangle &=  \frac{1}{2}\e^{-i\phi}\sin\theta .
\end{eqnarray}
\endnumparts
Hence,
\begin{equation}
\label{Qop}
\fl \langle s|{\mathbf Q}|s\rangle = Q(\theta,\phi)\equiv  -\beta \left (1-\e^{i\phi}\right )\frac{1+\cos\theta}{2} -\alpha \left (1-\e^{-i\phi}\right )\frac{1-\cos\theta}{2} .
\end{equation}

\subsection{CK equation}

Let us now return to the CK equation (\ref{CK}). Introduce the state vectors
\begin{equation}
|\psi_n(t)\rangle =\int_{-\infty}^{\infty} dx\, P_n(x,t) |x\rangle ,\quad n=0,1,
\end{equation}
and rewrite (\ref{CK}) in the operator form
\begin{eqnarray*}
\frac{d}{dt}|  \psi_n(t)\rangle&=\left [-i\hat{p}F_n(\hat{x})-\hat{p}^2 D_n(\hat{x})\right ]|\psi_n(t)\rangle +\sum_{m=0,1}Q_{nm}|\psi_m(t)\rangle.
\end{eqnarray*}
Set
\begin{equation}
\widehat{H}_n=-i\hat{p}F_n(\hat{x})-\hat{p}^2 D_n(\hat{x}),\quad n=0,1,
\end{equation}
and consider the diagonal matrix operator
\begin{equation}
\widehat{\bf K}=\left (\begin{array}{cc} \widehat{H}_0 &0 \\ 0&  \widehat{H}_1 \end{array} \right )= \left (\frac{1}{2}{\mathbf 1}  +\sigma_z\right) \widehat{H}_0+\left (\frac{1}{2}{\mathbf 1} -\sigma_z\right)  \widehat{H}_1 .
\end{equation}
We can thus rewrite (\ref{CK}) as an operator equation
\begin{eqnarray}
\label{psiH}
\frac{d}{dt}|  {\bm \psi}(t)\rangle&=\widehat{\bf H}|  {\bm \psi}(t)\rangle,\quad |  {\bm \psi}(t)\rangle=(|  {\psi}_0(t)\rangle,|  { \psi}_1(t)\rangle)^{\top},
\end{eqnarray} 
with
\begin{eqnarray}
\label{sh:H}
\fl \widehat{\bf H}&=\widehat{\bf K}+{\bf Q} \\
\fl &= \left (\frac{1}{2}{\mathbf 1}  +\sigma_z\right) \widehat{H}_0+\left (\frac{1}{2}{\mathbf 1} -\sigma_z\right)  \widehat{H}_1-\beta \left (\frac{1}{2}{\mathbf 1} +\sigma_z\right) -\alpha \left (\frac{1}{2}{\mathbf 1} -\sigma_z\right) +\alpha \sigma_++\beta\sigma_-.\nonumber
\end{eqnarray}
Moreover,
\begin{eqnarray}
\fl \langle s|\widehat{\bf H}|s\rangle&=H(\theta,\phi,\hat{x},\hat{p})\nonumber \\
\fl &\equiv  -\left (\beta \left [1-\e^{i\phi}\right ]-\widehat{H}_0\right )\frac{1+\cos\theta}{2} - \left (\alpha \left [1-\e^{-i\phi}\right ]-\widehat{H}_1\right )\frac{1-\cos\theta}{2} .
\label{HH}
\end{eqnarray}
Formally integrating equation (\ref{psiH}) gives
\begin{equation}
\label{sh:sol}
|{\bm \psi}(t)\rangle=\e^{\widehat{\bf H} t}|{\bm \psi}(0)\rangle.
\end{equation}

\subsection{Three-state model} 
\begin{figure}[h!]
\raggedleft
\includegraphics[width=9cm]{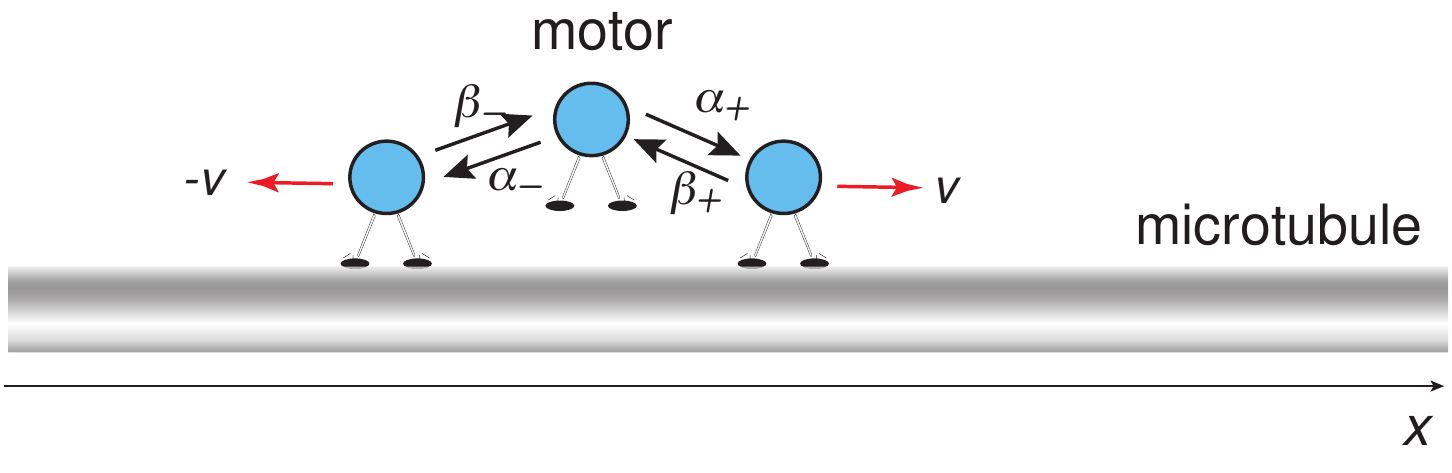}
\caption{Three-state motor transport model.}
\label{fig2}
\end{figure}

The above construction can be extended to three or more states, although the analysis becomes more complicated. Here we will restrict our discussion to a three-state model with $N(t)\in \{0,1,2\}$ and a matrix generator of the form 
\begin{equation}
\label{Q3}
{\bf Q}=\left (\begin{array}{ccc} -\beta_+ &\alpha_+ &0  \\ \beta_+ & -\alpha_+ -\alpha_- & \beta_- \\ 0 & \alpha_- &-\beta_- \end{array} \right ).
\end{equation}
This discrete process has recently appeared in a stochastic model of gene expression that includes three distinct histone states as well as two DNA promoter states \cite{Bhatt20}\footnote{Histones are proteins found in eukaryotic cell nuclei that pack and order the DNA into structural units called nucleosomes. They play an important role in epigenetics.}. Another example is the three-state model of motor transport shown in Fig. \ref{fig2}, which consists of a right-moving state ($n=0,v_0=v$), a stationary state ($n=1, v_1=0$), and a left-moving state ($n=2,v_2=-v$). Moreover, transitions can only occur either into or out of the stationary state.

We develop the coherent spin-$1$ state construction along the lines of \cite{Bhatt20}. First, rewrite the generator (\ref{Q3}) in the form
\begin{equation}
{\bf Q}=\alpha_+ {\bf T}_++\alpha_- {\bf T}_-+\beta_+ {\bf S}_++\beta_- {\bf S}_-,
\end{equation}
with
\begin{equation}
{\bf T}_+=\left (\begin{array}{ccc} 0& 1 &0 \\ 0& -1&0 \\ 0 &0 &0 \end{array} \right ),\quad{\bf T}_-=\left (\begin{array}{ccc} 0& 0 &0  \\ 0& -1&0 \\ 0 &1&0 \end{array} \right ),
\end{equation}
and
\begin{equation}
{\bf S}_+=\left (\begin{array}{ccc} -1& 0 &0 \\ 1& 0&0 \\ 0 &0 &0 \end{array} \right ),\quad{\bf S}_-=\left (\begin{array}{ccc} 0& 0 &0  \\ 0& 0&1 \\ 0 &0&-1 \end{array} \right ),
\end{equation}
Next we introduce the coherent spin-$1$ state 
\begin{equation}
|s\rangle = \left (\begin{array}{c} \e^{i\phi}\cos^4\theta/2  \\ 2\cos^2\theta/2\, \sin^2\theta/2 \\ \e^{-i\phi}\sin^4\theta/2\end{array} \right ),\quad 0\leq \theta \leq \pi,\ 0\leq \phi <2\pi,
\end{equation}
together with the adjoint
\begin{equation}
\langle s|=\left (  \e^{-i\phi}, \, 1,\, \e^{i\phi} \right ).
\end{equation}
Note that 
\begin{equation}
\fl \langle s'|s\rangle =\e^{i(\phi-\phi')}\cos^4\theta+2\cos^2\theta/2\, \sin^2\theta/2+\e^{-i(\phi-\phi')}\sin^4\theta/2,
\end{equation}
so that $\langle s|s\rangle =1$ and
\begin{equation}
\label{sdiff3}
\fl \langle s+\Delta s |s\rangle =1-i \Delta \phi (\cos^4 \theta/2-\sin^4\theta/2) +O(\Delta \phi^2)\approx 1-i\Delta \phi \cos \theta.
\end{equation}
We also have the completeness relation
\begin{equation}
\label{comp3}
\frac{3}{4\pi} \int_0^{\pi}\sin \theta \, d\theta \int_0^{2\pi}d\phi\, |s\rangle \langle s|=1.
\end{equation}

It can checked that the following identities hold:
\numparts
\begin{eqnarray}
\langle s|{\bf T}_+|s\rangle &=2(\e^{-i\phi}-1)\cos^2\theta/2\, \sin^2\theta/2, \\
\langle s|{\bf T}_-|s\rangle &=2(\e^{i\phi}-1)\cos^2\theta/2\, \sin^2\theta/2, \\
\langle s|{\bf S}_+|s\rangle &=  (\e^{i\phi}-1)\cos^4\theta/2,\\
\langle s|{\bf S}_-|s\rangle &=  (\e^{-i\phi}-1)\sin^4\theta/2.
\end{eqnarray}
\endnumparts
Hence,
\begin{eqnarray}
\label{Qop3}
\fl \langle s|{\mathbf Q}|s\rangle &= Q(\theta,\phi)\equiv  - \left (1-\e^{i\phi}\right )\left ( 2\alpha_-\cos^2\theta/2\, \sin^2\theta/2+\beta_+\cos^4\theta/2\right )\\
\fl &\quad - \left (1-\e^{-i\phi}\right )\left ( 2\alpha_+\cos^2\theta/2\, \sin^2\theta/2+\beta_-\sin^4\theta/2\right ).
\nonumber
\end{eqnarray}
Finally, the operator equation (\ref{psiH}) becomes
\begin{eqnarray}
\label{psiH3}
\frac{d}{dt}|  {\bm \psi}(t)\rangle&=\widehat{\bf H}|  {\bm \psi}(t)\rangle,\quad |  {\bm \psi}(t)\rangle=(|  {\psi}_0(t)\rangle,|  { \psi}_1(t)\rangle,|  { \psi}_2(t)\rangle)^{\top},
\end{eqnarray} 
with
\begin{eqnarray}
\fl \langle s|\widehat{\bf H}|s\rangle&=Q(\theta,\phi)+\widehat{H}_0\cos^4\theta/2 +2\widehat{H}_1\cos^2\theta/2\sin^2\theta/2+\widehat{H}_2\sin^4\theta/2,
\end{eqnarray}
and
\begin{equation}
\widehat{H}_n=-i\hat{p}F_n(\hat{x})-\hat{p}^2 D_n(\hat{x}),\quad n=0,1,2.
\end{equation}

\section{Construction of stochastic hybrid path integral}

One of the advantages of expressing the evolution equation for the probability density in terms of an operator equation acting on a Hilbert space is that it is relatively straightforward to construct a corresponding path integral representation of the solution. For simplicity, we consider the two-state model and then indicate how to extend the construction to the three-state model. As with other stochastic processes, the first step is to divide the time interval $[0,t]$ into $N$ subintervals of size $\Delta t=t/N$ and rewrite the formal solution (\ref{sh:sol}) as
\begin{equation}
|{\bm \psi}(t)\rangle =\e^{\widehat{\bf H}\Delta t}\e^{\widehat{\bf H}\Delta t}\cdots \e^{\widehat{\bf H}\Delta t}|{\bm \psi}(0)\rangle,
\end{equation}
with $\widehat{\bf H}$ given by equation (\ref{sh:H}). We then insert multiple copies of appropriately chosen completeness relations. Here we use the completeness relations (\ref{fpcom1}) and (\ref{comp2}), which are applied to the product Hilbert space with $|s,x\rangle =|s\rangle \otimes |x\rangle$. Introducing the solid angle integral
\begin{equation}
\int_{\Omega}ds =\frac{1}{2\pi} \int_0^{\pi}\sin \theta \, d\theta \int_0^{2\pi}d\phi,
\end{equation}
 we have
\begin{eqnarray}
& |{\bm \psi}(t)\rangle =\int_{\Omega}ds_0 \cdots \int_{\Omega}ds_N   \int_{-\infty}^{\infty}dx_0\cdots  \int_{-\infty}^{\infty}dx_N
|s_N,x_N\rangle \nonumber\\
&\times \langle s_N,x_N|\e^{\widehat{\bf H}\Delta t}|s_{N-1},x_{N-1}\rangle \langle s_{N-1},x_{N-1}|\e^{\widehat{\bf H}\Delta t}|s_{N-2},x_{N-2}\rangle\nonumber \\
 &\cdots \times \langle s_1,x_1|\e^{\widehat{\bf H}\Delta t}|s_0,x_0\rangle \langle s_0,x_0|{\bm \psi}(0)\rangle .
 \label{pip}
\end{eqnarray}

In the limit $N\rightarrow \infty$ and $\Delta t \rightarrow 0$ with $N\Delta t =t$ fixed, we can make the approximation
\begin{eqnarray}
\fl &\langle s_{j+1},x_{j+1}|\e^{\widehat{\bf H}\Delta t}| s_{j},x_{j}\rangle \approx \langle s_{j+1},x_{j+1}|1+\widehat{\bf H}\Delta t| s_{j},x_{j}\rangle\nonumber \\
\fl &=\langle s_{j+1} | s_{j} \rangle\bigg \{\delta(x_{j+1}-x_{j}) +\langle x_{j+1}|H(\theta_j,\phi_j,x_j,\hat{p}_j)\Delta t | x_j\rangle\bigg \}+O(\Delta t^2),
\end{eqnarray}
with $H$ defined in equation (\ref{HH}).
In addition, equation (\ref{sdiff}) implies that
\begin{eqnarray}
\fl \langle s_{j+1}|s_j\rangle = 1-\frac{1}{2}i(\phi_{j+1}-\phi_j)\cos \theta_j+O(\Delta \phi^2)=1-\frac{1}{2}i\Delta t\frac{d\phi_j}{dt}\cos\theta_j +O(\Delta t^2).
\end{eqnarray}
Each small-time propagator thus becomes (to first order in $\Delta t$)
\begin{eqnarray}
\label{K1}
 &\langle s_{j+1},x_{j+1}|\e^{\widehat{\bf H}\Delta t}| s_{j},x_{j}\rangle \\
 &\approx \langle x_{j+1}|\exp\left (\left [H(\theta_j,\phi_j,x_j,\hat{p}_j)-\frac{i}{2}\frac{d\phi_j}{dt}\cos\theta_j \right ]\Delta t\right )| x_{j}\rangle .\nonumber
\end{eqnarray}
If we now substitute the momentum completeness relation (\ref{fpcom2}) into the small-time propagator (\ref{K1}) we see that
\begin{eqnarray}
\fl &\langle s_{j+1},x_{j+1}|\e^{\widehat{\bf H}\Delta t}| s_{j},x_{j}\rangle \nonumber \\
\fl &\approx \int_{-\infty}^{\infty}\frac{dp_j }{2\pi}\langle x_{j+1}| p_j\rangle  \langle p_j|x_j\rangle  \exp\left (\left [H(\theta_j,\phi_j,x_j,p_j)-\frac{i}{2}\frac{d\phi_j}{dt}\cos\theta_j \right ]\Delta t\right ).
\label{K2}
\end{eqnarray}
Furthermore,
\begin{eqnarray}
\label{K3}
\langle x_{j+1}| p_j\rangle  \langle p_j|x_j\rangle =\e^{ip_j(x_{j+1}-x_j)}=\exp\left (ip_j\frac{dx_j}{dt}\Delta t\right )+O(\Delta t^2).
\end{eqnarray}
Substituting equations (\ref{K2}) and (\ref{K3}) into (\ref{pip}) yields 
\begin{eqnarray}
\fl& |{\bm \psi}(t)\rangle =\int_{\Omega}ds_0 \cdots \int_{\Omega}ds_N  \int_{-\infty}^{\infty}\int_{-\infty}^{\infty}\frac{dx_0dp_0}{2\pi}\cdots  \int_{-\infty}^{\infty}\int_{-\infty}^{\infty}\frac{dx_{N}dp_{N}}{2\pi}  
|s_N,x_N\rangle \nonumber\\
\fl &\quad \times \prod_{j=0}^{N-1} \exp\left (\left [H(\theta_j,\phi_j,x_j,p_j)-\frac{i}{2}\frac{d\phi_j}{dt}\cos\theta_j +ip_j\frac{dx_j}{dt}\right ]\Delta t\right )\langle s_0,x_0|{\bm \psi}(0)\rangle . \nonumber
 \label{pip2}
\end{eqnarray}
The final step is to take the continuum limit $N\rightarrow \infty,\Delta t\rightarrow 0$ with $N\Delta t=t$ fixed, $x_j=x(j\Delta t)$ etc. We will also assume that
\[<x_0|\psi_n(0)\rangle =\rho_n \delta(x-x_0),
\]
and set
\[P_n(x,t|x_0,0)=\langle x,n|{\bm \psi}(t)\rangle.\]
After Wick ordering, $p\rightarrow -ip$, integrating by parts the term involving $d\phi/dt$, and performing the change of coordinates $z=(1+\cos \theta)/2$, we obtain the following functional path integral:
\begin{eqnarray}
\fl  P_n(x,t|x_0,0)&={\mathcal N}_n \int_{x(0)=x_0}^{x(t)=x} {\mathcal D}[\phi]{\mathcal D}[z]  {\mathcal D}[p]{\mathcal D}[x] \exp\left (-\int_{0}^{t}\left [p\frac{dx}{d\tau}-i\phi\frac{dz}{d\tau} -{\mathcal H}\right ]d\tau\right ), \nonumber \\
\fl
\label{pieff}
\end{eqnarray}
where ${\mathcal N}_n$ is a constant and ${\mathcal H}$ is the effective ``Hamiltonian'' 
\begin{eqnarray}
 {\mathcal H}&= \left (-\beta \left [1-\e^{i\phi}\right ] +pF_0(x)+p^2 D_0(x)\right )z\nonumber\\
&\quad + \left (-\alpha \left [1-\e^{-i\phi}\right ]+pF_1(x)+p^2 D_1(x)\right )(1-z),
\label{Heff}
\end{eqnarray}
with ``position'' coordinates $(x,z)$ and ``conjugate'' momenta $(p,-i\phi)$. It will turn out that $z(t)$ represents the probability that the discrete state $N(t)=0$ at time $t$.
\medskip

\subsection{Some remarks}

\noindent (i) If $F$ and $D$ are independent of the discrete state $n$ and the transition rates $\alpha,\beta$ are $x$-independent, then the path integral (\ref{pip2}) reduces to the product of two independent path integrals, corresponding to the continuous and discrete processes, respectively:
\begin{eqnarray}
\fl & P_n(x,t|x_0,0)=\int \limits_{x(0)=x_0}^{x(t)=x}  {\mathcal D}[p]{\mathcal D}[x]\exp\left (-{\int_{0}^{t}[p\dot{x}-pF(x)-p^2D(x)]d\tau}\right )\\
\fl &\quad  \times {\mathcal N}_n \int {\mathcal D}[\phi]{\mathcal D}[z]\exp\left (-\int_{0}^{t}\left [-i\phi \dot{z}+\beta \left [1+\e^{i\phi}\right ]z +\alpha \left [1+\e^{-i\phi}\right ](1-z) \right ]d\tau\right ).\nonumber
\end{eqnarray}
That is, the path integral of the continuous stochastic process decouples from the discrete process and we recover the standard action of a one-dimensional Ito SDE \cite{Martin73,Dom76,Janssen76,Holmes20}:
\begin{equation}
\label{Ssde}
S[x,p]=\int_{0}^{t}[p\dot{x}-pF(x)-p^2D(x)]d\tau.
\end{equation}

\medskip

\noindent (iii) The derivation carries over to higher-dimensional stochastic hybrid systems with $M$ continuous variables $x_{\ell}$, $\ell=1,\ldots M$. The  Ito SDE becomes
\begin{equation}
\label{pdmulti}
dX_{\ell}=F_{n,{\ell}}(\x)dt+\sqrt{2D_{n,\ell}(\x)}dW_{\ell}
\end{equation}
for $N(t)=n$, where $W_{\ell}(t)$ are independent Wiener processes. The multivariate CK equation takes the form
\begin{eqnarray}
\fl \frac{\partial P_n}{\partial t}&=\sum_{{\ell}=1}^M\left [-\frac{\partial}{\partial x_{\ell}}(F_{n,\ell}(\x)P_n(\x,t)) +\frac{\partial^2}{\partial x_{\ell}^2}(D_{n,\ell}(\x)P_n(\x,t))\right ]+\sum_{m}Q_{nm}P_m(\x,t).\nonumber \\
\fl
\label{CK0}
\end{eqnarray}
Following along identical lines to the one-dimensional case, one obtains
a path-integral representation of the solution to equation (\ref{CK0}):
\begin{eqnarray}
 P_n(\x,t|\x_0,0)&={\mathcal N}_n \int_{\x(0)=\x_0}^{\x(t)=\x} {\mathcal D}[\phi]{\mathcal D}[z]  {\mathcal D}[\p]{\mathcal D}[\x] \nonumber \\
&\quad \times \exp\left (-\int_{0}^{t}\left [\sum_{\ell=1}^M p_{\ell} \dot{x}_{\ell}-i\phi\dot{z}-{\mathcal H}\right ]d\tau\right ), 
\label{pieff2}
\end{eqnarray}
with
\begin{eqnarray}
 {\mathcal H}&= \left (-\beta \left [1-\e^{i\phi}\right ] +\sum_{\ell=1}^M [p_{\ell}F_{0,\ell}(\x)+p_{\ell}^2 D_{0,\ell}(\x)]\right )z\nonumber\\
&\quad + \left (-\alpha\left [1-\e^{-i\phi}\right ]+\sum_{\ell=1}^M [p_{\ell}F_{1,\ell}(\x)+p_{\ell}^2 D_{1,\ell}(\x)]\right )(1-z).
\label{Heff2}
\end{eqnarray}

\noindent (iii) The path integral construction can also be extended to the case of more than two discrete states. In particular, consider the three-state model of section 3.4. All of the steps in the derivation proceed as before. The final result is a path integral of the form (\ref{pieff}) with $dz/dt\rightarrow 2dz/dt$ and the modified Hamiltonian
\begin{eqnarray}
\label{Heff3}
\fl {\mathcal H} &=    \left (-\beta_+\left [1-\e^{i\phi}\right ]+pF_0(x)+p^2D_0(x) \right )z^2\\
\fl & \qquad +2\left (-\alpha_-\left [1-\e^{i\phi}\right ]-\alpha_+\left [1-\e^{-i\phi}\right ]+pF_1(x)+p^2D_1(x) \right )z(1-z)\nonumber
\\
\fl & \qquad +\left (-\beta_-\left [1-\e^{-i\phi}\right ]+pF_2(x)+p^2D_2(x) \right )(1-z)^2.\nonumber
\end{eqnarray}

\subsection{Semi-classical limit}

In general it is not possible to evaluate a stochastic path integral without some form of approximation scheme. One of the best known is the so-called semi-classical approximation, which involves expanding the path integral action to second order in the variables $p,\phi$, assuming that the system operates in the weak noise regime. In the case of the piecewise SDE (\ref{pd}), we define the weak noise limit by introducing the scalings $\alpha \rightarrow \alpha/\epsilon,\beta\rightarrow \beta/\epsilon$ and $D\rightarrow \epsilon D$. The former represents fast switching between the discrete states (adiabatic limit), whereas the latter represents weak Gaussian noise. Introducing the additional scaling $\phi\rightarrow \epsilon \phi$, the path integral (\ref{pieff}) for the two-state model becomes
\begin{eqnarray}
\fl  P_n(x,t|x_0,0)&={\mathcal N}_n \int_{x(0)=x_0}^{x(t)=x} {\mathcal D}[\phi]{\mathcal D}[z]  {\mathcal D}[p]{\mathcal D}[x] \e^{-S},
\label{pieg}
\end{eqnarray}
with the action
\begin{eqnarray}
S&=\int_{0}^{t}\bigg [p\dot{x}-i\epsilon \phi\dot{z} +\left (\frac{\beta}{\epsilon} \left [1-\e^{i\epsilon \phi}\right ] -  pF_0(x)-\epsilon p^2 D_0(x)\right )z\nonumber\\
&\quad + \left (\frac{\alpha}{\epsilon} \left [1-\e^{-i\epsilon \phi}\right ]-  pF_1(x)-\epsilon p^2 D_1(x)\right )(1-z)\bigg ]d\tau .
\label{act0}
\end{eqnarray}
Under the approximation $1-\e^{\pm i \epsilon\phi}=\mp i\epsilon \phi+\epsilon^2\phi^2/2+\ldots$, the action is quadratic in $p,\phi$. Comparison with the action (\ref{Ssde}) of an SDE then establishes that the resulting path integral represents the probability density functional of an effective stochastic processes evolving according to the following pair of coupled Langevin equations:
\numparts
\begin{eqnarray}
\label{xdot}
\frac{dx}{dt}&=F_0(x)z+F_1(x)(1-z)+\sqrt{2\epsilon}\xi_x,\\
 {\epsilon}\frac{dz}{dt}&=-\beta z+\alpha(1-z)+\sqrt{2\epsilon}\xi_z,
 \label{zdot}
\end{eqnarray}
\endnumparts
where $\xi_x$ and $\xi_z$ are independent Gaussian white noise processes with $\langle \xi_x\rangle =0=\langle \xi_z\rangle$ and
\numparts
\begin{eqnarray}
\label{Lan1}
\langle \xi_x(t) \xi_x(t')\rangle&=   [D_0(x)z+D_1(x)(1-z)] \delta(t-t')  ,          \\
\langle \xi_z(t)\xi_z(t')\rangle&=  \frac{1}{2}(\beta z+\alpha(1-z))\delta(t-t').
\label{Lan2}
\end{eqnarray}
\endnumparts
Note that the multiplicative noise term $\xi_z$ in equation (\ref{zdot}) vanishes at $z=0,1$, ensuring that the stochastic variable $z(t)$ remains within the domain $[0,1]$. Therefore, we can interpret $z(t)$ as an auxiliary variable that represents the effective probability that $N(t)=0$ at time $t$. An analogous result holds for the 3-state model of section 3.4. That is, $z(t)$ parameterizes an effective probability distribution $\psi_n$, $n=0,1,2$, with $\psi_0=z^2$, $\psi_1=2z(1-z)$ and $\psi_3=(1-z)^2$. (More generally, for $N$ discrete states $\psi_n$ is generated by considering the binomial expansion of $(z+(1-z))^N$.)

A further approximation can be obtained by using a linear noise approximation. Taking the limit $\epsilon \rightarrow 0$ shows that $z(t)\rightarrow z^*=\alpha/(\alpha+\beta)$ and $x(t)$ satisfies the deterministic mean-field equation
\begin{equation}
\frac{dx}{dt}=\overline{F}(x)=F_0(x)z^*(x)+F_1(x)(1-z^*(x)),
\end{equation}
assuming $x$-dependent transition rates. Clearly $\rho_0=z^*$ and $\rho_1=1-z^*$ is the stationary distribution of the two-state Markov chain (\ref{mc}).
Substituting $z(t)=z^*+y(t)$ into equation (\ref{Lan2}) implies that to leading order,
\[y(t)=\frac{\sqrt{2\epsilon}}{\alpha+\beta}\xi_0(t),\]
with
\[\langle \xi_0(t)\rangle =0,\quad \langle \xi_0(t)\xi_0(t')\rangle = \beta z^* \delta(t-t').\]
Applying the linear noise approximation to equation (\ref{Lan1}) then gives
\begin{equation}
\label{lna1}
\fl \frac{dx}{dt}=\overline{F}(x)+[F_0(x)-F_1(x)]y+\sqrt{2\epsilon}\xi_x=\overline{F}(x)+\sqrt{2\epsilon}\xi+\sqrt{2\epsilon}\xi_x,
\end{equation}
where
\begin{equation}
\label{lna2}
\fl \langle \xi(t)\rangle =0,\quad \langle \xi(t)\xi(t')\rangle =D_{\rm eff}(x)\delta(t-t'),\quad D_{\rm eff}(x)\equiv \frac{\alpha\beta}{\alpha+\beta}\frac{[F_0(x)-F_1(x)]^2}{(\alpha+\beta)^2} .
\end{equation}
A little algebra shows that
\[D_{\rm eff}(x)=\frac{\alpha[F_0(x)-\overline{F}(x)]^2+\beta[F_1(x)-\overline{F}(x)]^2}{(\alpha+\beta)^2},
\]
which is precisely the effective diffusion coefficient obtained using a quasi-steady-state approximation of the CK equation 
\begin{eqnarray}
\label{CK00}
\frac{\partial P_n(x,t)}{\partial t}=-\frac{\partial F_n(x)P_n(x,t)}{\partial x}+\frac{1}{\epsilon}\sum_{m= 0,1}Q_{nm}(x)P_m(x,t),
\end{eqnarray}
in the fast switching limit \cite{Newby10}. The latter method is based on substituting the solution $P_n(x,t)=C(x,t)\rho_n +\epsilon w_n(x,t)$ into the CK equation and deriving a Fokker-Planck equation for $C(x,t)$ using the Liapunov-Schmidt procedure. It essentially assumes that $z(t)\approx z^*$ as in the linear noise approximation.
One of the nice features of the path integral representation based on coherent spin states is that it also keeps track of $z(t)$. In particular, one can explore the stochastic dynamics under the semiclassical approximation in the weakly nonadiabatic regime, by considering the coupled system of Langevin equations (\ref{Lan1}) and (\ref{Lan2}). This observation has also been made within the specific context of stochastic gene expression \cite{Zhang13}. 

\subsection{Example}

As a simple illustration of the semi-classical limit, consider a two-state gene network without feedback, see equation (\ref{autoreg}) and Fig. \ref{fig2}.  (In the limit $\gamma \rightarrow 0$ this model is equivalent to the motor transport model with $v_0=\kappa$ and $v_1=0$.)
The pair of Langevin equations takes the form (after setting $z=z^*+y$)
\begin{eqnarray}\label{gene}
\frac{dx}{dt}&=\kappa z^*-\gamma x+\kappa y,\quad 
 {\epsilon}\frac{dy}{dt}&=-(\alpha+\beta) y+\sqrt{2\epsilon}\xi_y,
\end{eqnarray}
with $\alpha,\beta$ constant, $z^*=\alpha /(\alpha+\beta)$ and
\begin{eqnarray} 
\langle \xi_y(t)\xi_y(t')\rangle&=  \frac{1}{2}(2\beta z^*+(\beta-\alpha)y)\delta(t-t').
\end{eqnarray}
Integrating the equations (\ref{gene}) with respect to $t$ yields
\begin{equation}
{x}(t)={x}_0\e^{-\gamma t}+\frac{\kappa z^*}{\gamma}\left (1-\e^{-\gamma t}\right )+\kappa \int_0^t
\e^{-\gamma (t-t')}y(t')dt',
\label{xgene}
\end{equation}
and
\begin{equation}
{y}(t)={y}_0\e^{-(\alpha+\beta) t/\epsilon }+\sqrt{\frac{2}{\epsilon}}\int_0^t
\e^{-(\alpha+\beta) (t-t')/\epsilon}\xi_y(t')dt'.
\label{ygene}
\end{equation}
Taking expectations of these two equations and substituting for $\bar{y}(t)$ into (\ref{xgene}), we obtain the following equation for the mean protein concentration $\bar{x}$:
\begin{equation}
\fl \bar{x}(t)={x}_0\e^{-\gamma t}+\frac{\kappa z^*}{\gamma}\left (1-\e^{-\gamma t}\right )+\frac{\epsilon \kappa y_0}{(\alpha+\beta)-\epsilon \gamma}
\left (\e^{-\gamma t} -\e^{-(\alpha+\beta)t/\epsilon}\right ).
\end{equation}
It follows that taking into account the dynamics of the auxiliary variable $z(t)$ leads to additional contributions to the dynamics of the mean protein concentration that are missed by the mean field equation. A similar result holds for the variance, which is given by
\begin{eqnarray}
 &\langle [x(T)-\bar{x}(T)]^2\rangle=2\kappa^2\int_0^T\int_0^t
\e^{-\gamma (T-t)}\e^{-\gamma (T-t')}\Delta_y(t,t') dt'dt,
\end{eqnarray}
where (for $t'<t$)
\begin{eqnarray}
\fl \Delta_y(t,t')&\equiv \langle [y(t)-\bar{y}(t)][y(t')-\bar{y}(t')]\rangle \nonumber \\
\fl & ={\frac{2}{\epsilon}}\int_0^t
\e^{-\Gamma(t-\tau)}\int_0^{t'}
\e^{-\Gamma (t'-\tau') }\langle \xi_y(\tau)\xi_y(\tau')\rangle d\tau'd\tau\nonumber \\
\fl &=\frac{1}{\epsilon}\int_0^{t'} \e^{-\Gamma (t+t'-2\tau)}(2\beta z^*+(\beta-\alpha))\bar{y}(\tau)d\tau\nonumber \\
\fl &=\frac{\beta z^*}{\alpha+\beta}\left [ \e^{-\Gamma (t-t')}-\e^{-\Gamma (t+t')}\right ]+\frac{ (\beta-\alpha)y_0}{(\alpha+\beta)}\left (\e^{-\Gamma t }-\e^{-\Gamma (t+t')}\right ),
\end{eqnarray}
After some algebra we find that
\begin{eqnarray}
\fl  &\langle [x(T)-\bar{x}(T)]^2\rangle= \frac{2\kappa^2 \e^{-2\gamma T}}{\alpha+\beta} \bigg \{\frac{\beta z^*}{\gamma+\Gamma} \left [\frac{\e^{2\gamma T}-1}{2\gamma}-\frac{\e^{(\gamma -\Gamma)T}-1}{\gamma-\Gamma}\right ]\nonumber \\
\fl &\quad -\frac{\beta z^* }{\gamma-\Gamma}\left [\frac{\e^{2(\gamma -\Gamma)T}-1}{2(\gamma-\Gamma)}-\frac{\e^{(\gamma -\Gamma)T}-1}{\gamma-\Gamma}\right ]+\frac{(\beta -\alpha)y_0}{\gamma} \left [\frac{\e^{(2\gamma-\Gamma) T}-1}{2\gamma-\Gamma}-\frac{\e^{(\gamma -\Gamma)T}-1}{\gamma-\Gamma}\right ]\nonumber\\
\fl &\quad -\frac{(\beta -\alpha)y_0}{\gamma-\Gamma} \left [\frac{\e^{2(\gamma-\Gamma) T}-1}{2(\gamma-\Gamma)}-\frac{\e^{(\gamma -\Gamma)T}-1}{\gamma-\Gamma}\right ]\bigg \}.
\label{dood}
\end{eqnarray}
In the limit $\epsilon \rightarrow 0$, we can drop all exponentially small terms $\e^{-\Gamma T}=\e^{-(\alpha+\beta)T/\epsilon}$. The remaining terms generate a power series in $\epsilon$ whose leading order form is
\begin{eqnarray*}
 \langle [x(T)-\bar{x}(T)]^2\rangle&= \frac{2 \kappa^2 }{\alpha+\beta} \frac{\epsilon \beta z^*}{\alpha+\beta}   \frac{1- \e^{-2\gamma T}}{2\gamma}  +O(\epsilon^2).\end{eqnarray*}
 The $O(\epsilon)$ terms is identical to the variance obtained  from the linear noise approximation of equations (\ref{lna1}) and (\ref{lna2}), which become
\begin{equation}
\label{xdot2}
\frac{dx}{dt}=\kappa z^*-\gamma x+\sqrt{2\epsilon}\xi,
\end{equation}
with
\begin{equation}
\langle \xi(t)\rangle =0,\quad \langle \xi(t)\xi(t')\rangle =D \delta(t-t'),\quad D \equiv \frac{\alpha\beta}{\alpha+\beta}\frac{\kappa^2}{(\alpha+\beta)^2} .
\end{equation}
(Note that certain care has to be taken in interpreting equation (\ref{xdot2}), since there is a small but non-zero probability that the concentration $x(t)$ can become negative.)

\section{Alternative path integral representation and least action paths}

The form of the path integral (\ref{pieff}) differs significantly from a previous version that was derived in the weak noise limit using either integral representations of the Dirac delta function \cite{Bressloff13a,Bressloff15} or operator methods adapted from Doi-Peliti \cite{Bressloff21}. One of the major applications of the second version is that it provides a relatively straightforward method for calculating least-action paths in noise-induced escape problems \cite{Bressloff14a}. 

For the moment, let us consider the hybrid system (\ref{pd}) with an arbitrary number of discrete states $n=0,1,\ldots,N$ evolving according to an irreducible Markov chain with generator ${\bf Q}$. The discrete process is said to be irreducible if there exists a $t>0$ such that $\e^{{\bf Q}t}>0$; this implies that any two states of the Markov chain can be connected in a finite time. One can then apply the Perron-Frobenius theorem for finite square matrices \cite{Grimmett}. In particular, there exists a unique positive right-eigenvector $\rho_n$ for which $\sum_{m}Q_{nm}\rho_m=0$; the corresponding left eigenvector is $(1,1,\ldots,1)$ since $\sum_{n}Q_{nm}=0$. We can identify $\rho$ as the unique stationary density. Moreover, the Perron Frobenius theorem ensures that all other eigenvalues have negative real parts, ensuring that the distribution $P_m(t)\rightarrow \rho_m$ as $t\rightarrow \infty$. Now consider the generalized eigenvalue equations
\numparts
\begin{eqnarray}
\label{eigR}
\fl &\sum_{m=1}^N\left \{ [pF_n(x)+p^2 D_n(x)]\delta_{m,n} +Q_{nm}(x)\right \}R_{m}(x,p) =\Lambda(x,p)R_{n}(x,p)  ,  \\
\fl &\sum_{m\geq 0}\overline{R}_{m}(x,p)\left \{[pF_n(x)+p^2 D_n(x)]\delta_{m,n} +Q_{mn}(x)\right \}=\Lambda(x,p)\overline{R}_{n}(x,p),\label{eigS}
\end{eqnarray} 
\endnumparts
with
\begin{equation}
\label{RS}
\sum_{m=1}^NR_{m}(x,p)\overline{R}_{m}(x,p)=1.
\end{equation}
Note that when $p=0$ we recover the eigenvalue equation for ${\bf Q}$. The Perron-Frobenius theorem can also be applied to the linear system (\ref{eigR}), which means that there exists a unique principal eigenvalue $\Lambda(x,p)$ and associated positive eigenvector ${\bf R}(x,p)$.
In the weak noise limit (as defined above), one obtains the following path integral representation of the solution to the corresponding CK equation \cite{Bressloff13a,Bressloff15,Bressloff21}:
\begin{eqnarray}
\label{piold}
\fl  & P_n(x,t)={\mathcal N}_n\int \limits_{x(0)=x_0}^{x(t)=x}  {\mathcal D}[p]{\mathcal D}[x] \exp\left (-\frac{1}{\epsilon}{\int_{0}^{t}[p\dot{x}-\Lambda(x,p)]d\tau}\right ). 
\end{eqnarray}
The principal eigenvalue $\Lambda$ acts as an effective Hamiltonian with the parameter $p$ of the eigenvalue equation (\ref{eigR}) playing the role of a momentum variable. (The resulting classical action can also be derived without the use of path integrals \cite{Bressloff17} using  the variational LDP for hybrid systems introduced by Faggionato et al. \cite{fagg09,Faggionato10}; a rigorous but rather technical derivation can be found in \cite{Kifer09}.)

In the limit $\epsilon \rightarrow 0$, the path integral is dominated by least-action paths, which satisfy Hamilton's equations
\begin{equation}
  \dot{x} = \frac{\partial \Lambda(x,p)}{\partial p}, \quad  \dot{p} =- \frac{\partial \Lambda(x,p)}{\partial x},
\end{equation}
Differentiating both sides of equation (\ref{eigR}) with respect to $p$ gives
\begin{eqnarray*}
\fl &\sum_{m=1}^N\left \{ [F_n +2p D_n ]\delta_{m,n} +Q_{nm} \right \}R_{m} +\sum_{m=1}^N\left \{ [pF_n +p^2 D_n ]\delta_{m,n} +Q_{nm} \right \}\frac{\partial R_{m} }{\partial p}  \\
\fl &\qquad  =\Lambda \frac{\partial R_{n} }{\partial p}+ \frac{\partial \Lambda }{\partial p}R_n .
\end{eqnarray*} 
Summing over $n$ with $\sum_nQ_{nm}=0$ and setting $p=0$ thus shows that
\[ \left . \frac{\partial \Lambda }{\partial p}\right |_{p=0}=\sum_n F_n\rho_n =\overline{F},\]
and we recover the deterministic mean-field equation $\dot{x}=\overline{F}(x)$. (It can also be checked that $\dot{p}=0$ at $p=0$.) However, there also exist least action paths for which $p\neq0$. In particular, the zero energy paths with $\Lambda(x,p)=0$ represent the most likely paths of escape from a metastable state $x^*$. Evaluating the action along such a path yields the so-called quasipotential
\begin{equation}
\Phi_0(x)=\int_{-\infty}^Tp(t)\dot{x}(t)dt,
\end{equation}
with $x(T)=x$ and $x(-\infty)=x^*$. The latter is also the solution of the Hamilton-Jacobi equation
\begin{equation}
\label{HJ}
\lambda(x,\Phi_0'(x))=0.
\end{equation}
Combining the evaluation of least-action paths with matched asymptotic methods provides an estimate for the mean first passage time to escape from a metastable state, which has the exponential form $\tau\sim \e^{\Phi_0(x)/\epsilon}$ \cite{Bressloff14a}.

In the case of the two-state hybrid model (\ref{pd}) it is possible to determine $\Lambda$ explicitly. The linear equation (\ref{eigR}) can be written as the two-dimensional system
\begin{equation}
 \left (\begin{array}{cc} -\beta+pF_0(x)& \alpha \\ \beta & -\alpha+pF_1(x) \end{array}\right )\left (\begin{array}{c} R_0 \\ R_1\end{array}\right )=\Lambda \left (\begin{array}{c} R_0 \\ R_1\end{array}\right ).
\end{equation}
Solving the corresponding characteristic equation yields the
principle eigenvalue 
\begin{eqnarray}
\Lambda(x,p)=\frac{1}{2}\left [  \Sigma(x,p)+\sqrt{\Sigma(x,p)^2- 4\Gamma(x,p)}    \right ],
\end{eqnarray}
where
\[\Sigma(x,p)=p(F_0(x)+F_1(x))-[\alpha+\beta],\]
and
\[\Gamma(x,p)=(pF_1(x)-\alpha)(pF_0(x)-\beta)-\alpha \beta.\]
A little algebra shows that
\[ {\mathcal A}(x,p)\equiv \Sigma(x,p)^2- 4\Gamma(x,p)=[p(F_0-F_1)-(\alpha+\beta)]^2+\alpha\beta >0,\]
so that as expected $\Lambda$ is real. From Hamilton's equations
\begin{eqnarray}
\fl  \dot{x}&=&\frac{\partial \Lambda(x,p)}{\partial p} =\frac{F_0(x)+F_1(x)}{2}+\frac{\partial {\mathcal A}(x,p)}{\partial p}\frac{1}{2\sqrt{{\mathcal A}(x,p)}} \nonumber \\
 \fl &=&\frac{F_0(x)+F_1(x)}{2}+\frac{F_0(x)-F_1(x)}{2}\frac{p(F_0-F_1)-(\alpha-\beta)}{\sqrt{[p(F_0-F_1)+(\alpha-\beta)]^2+\alpha\beta }}.
\end{eqnarray}
Moreover,
writing
\[\dot{x}=F_0(x)\psi_0(x,p)+F_1(x)\psi_1(x,p),\]
we see that
\begin{equation}
\psi_0(x,p)=\frac{1}{2}\left [1+\frac{p(F_0-F_1)+(\alpha-\beta)}{\sqrt{[p(F_0-F_1)-(\alpha-\beta))]^2+\alpha\beta }}
\right ],
\end{equation}
and
\begin{equation}
\psi_1(x,p)=\frac{1}{2}\left [1-\frac{p(F_0-F_1)+(\alpha-\beta)}{\sqrt{[p(F_0-F_1)-(\alpha-\beta)]^2+\alpha\beta }}\right ],
\end{equation}
so that $\psi_{0,1} \geq 0$ with $\psi_0+\psi_1=1$. This suggests that $\psi_0$ plays an analogous role to the dynamical variable $z(t)$ in the path integral (\ref{pieff}).

\section{Equivalence of path integral representations in the weak noise limit}

In this section we show how the path integral representation (\ref{pieff}) reduces to the path integral (\ref{piold}) in the weak noise limit.
At first sight it is not clear how the principal eigenvalue $\Lambda(x,p)$ emerges from (\ref{pieff}) for $\epsilon \rightarrow 0$. Indeed, one has to go beyond the semi-classical limit considered in section 4. We begin by considering the two state model with Hamiltonian (\ref{Heff}). As a first step, consider the scalings $\alpha \rightarrow \alpha/\epsilon,\beta\rightarrow \beta/\epsilon$, $D\rightarrow \epsilon D$ and $p\rightarrow p/\epsilon$ and rewrite the path integral (\ref{pieff}) as
\begin{eqnarray}
\fl  P_n(x,t|x_0,0)&={\mathcal N}_n \int_{x(0)=x_0}^{x(t)=x} {\mathcal D}[\phi]{\mathcal D}[z]  {\mathcal D}[p]{\mathcal D}[x] \e^{-S/\epsilon},
\label{piegw}
\end{eqnarray}
with the action
\begin{eqnarray}
\fl S&=\int_{0}^{t}\bigg [p\dot{x}-i \epsilon \phi\dot{z} + \beta  \left [1-\e^{i \phi}\right ] z+\alpha \left [1-\e^{-i  \phi}\right ](1-z)-h(x,p,z)\bigg ]d\tau ,
\label{act2}
\end{eqnarray}
and
\begin{eqnarray}
h(x,p,z)&=(pF_0(x)+p^2D_0(x) )z+(pF_1(x)+p^2D_1(x) )(1-z).
\end{eqnarray}
(In contrast to equation (\ref{act0}), we have not rescaled $\phi$.) Next we define the functions
\begin{eqnarray}
q&=\beta_+ \left [1-\e^{i \phi}\right ] +\lambda ,\quad \lambda=- \alpha\left [1-\e^{-i\phi}\right ].
\end{eqnarray}
and rewrite the action as
\begin{eqnarray}
 S&=\int_{0}^{t}\bigg [p\dot{x}-i \epsilon \phi \dot{z} +  q(\phi)z  -\lambda(\phi) -h(x,p,z)\bigg ]d\tau .
 \label{act3}
\end{eqnarray}
Note that $\lambda$ and $q$ are related according to
\begin{equation}
\label{qlam}
q-\beta +\frac{\alpha\beta}{\lambda +\alpha}- \lambda =0.
\end{equation}
Moreover, equation (\ref{qlam}) is the characteristic equation for the eigenvalue equation
\begin{eqnarray}
&(q-\beta)r_0+\alpha r_1=\lambda r_0,\quad
\beta r_0-\alpha r_1=\lambda r_1.
\label{ooo}
\end{eqnarray}

In the limit $\epsilon \rightarrow 0$ the path integral will be dominated by paths that satisfy
\begin{equation}
\label{goog}
\epsilon \frac{dz}{dt}=\frac{dq}{d\phi}z-\frac{d\lambda}{d\phi}.
\end{equation}
This allows us to set $dz/dt=0$ in the action (\ref{act2}) and eliminate the independent variable $\phi$ by requiring
\begin{equation}
\frac{dq}{d\phi}z-\frac{d\lambda}{d\phi}=0,
\end{equation}
that is,
\begin{equation}
\label{zoo}
\e^{2i\phi}=\frac{\alpha (1-z)}{\beta z}.
\end{equation}
We thus obtain the reduced path integral
\begin{eqnarray}
 P_n(x,t|x_0,0)&={\mathcal N}_n \int_{x(0)=x_0}^{x(t)=x} {\mathcal D}[z]  {\mathcal D}[p]{\mathcal D}[x] \e^{-\widehat{S}/\epsilon},
\label{pired}
\end{eqnarray}
with
\begin{eqnarray}
 \widehat{S}&\approx \int_{0}^{t}\bigg [p\dot{x}+  {qz}  -\lambda -h(x,p,z)\bigg ]d\tau . 
\label{act33}
\end{eqnarray}
We can now eliminate $z$ by functionally minimizing the action $\widehat{S}$ with respect to $z$, noting that $q$ and $\lambda$ are functions of $z$ via their dependence on $\phi$:
\begin{eqnarray}
\label{dude}
 \fl 0&=\frac{\delta \widehat{S}}{\delta z(t)}= \frac{d q}{d\phi}\frac{d\phi}{dz}z+q-\frac{d \lambda}{d \phi }\frac{d \phi}{d z} -p[F_0(x)-F_1(x)]-p^2[D_0(x)-D_1(x)].
\end{eqnarray}
It then follows from equation (\ref{goog}) that
\begin{equation}
q=p[F_0(x)-F_1(x)]+p^2[D_0(x)-D_1(x)].
\end{equation}
Hence, the minimized action becomes
\begin{eqnarray}
\widehat{S}&= \int_{0}^{t}\bigg [p\frac{dx}{dt}   -\lambda  -  \left (pF_1(x)+  p^2 D_1(x)\right )\bigg ]d\tau .
\label{act4}
\end{eqnarray}
Finally, defining $\Lambda=\lambda +pF_1+p^2 D_1$ and substituting for $q$ and $\lambda$ in equations (\ref{ooo}) recovers the eigenvalue equation (\ref{eigR}), and hence (\ref{pired}) is equivalent to the path integral (\ref{piold}).

A similar reduction can be carried out for the three-state model with Hamiltonian (\ref{Heff3}). The action (\ref{3act2}) becomes
\begin{eqnarray}
\fl S&=\int_{0}^{t}\bigg [p\frac{dx}{dt}-2i \epsilon \phi\frac{dz}{dt} + \left (\beta_+\left [1-\e^{i\phi}\right ] \right )z^2\ + 2\left (\alpha_-\left [1-\e^{i\phi}\right ]+\alpha_+\left [1-\e^{-i\phi}\right ]\right )z(1-z)\nonumber \\
\fl &\quad +\left (\beta_-\left [1-\e^{-i\phi}\right ]\right )(1-z)^2-h(x,p,z)\bigg ]d\tau ,
\label{3act2}
\end{eqnarray}
where
\begin{eqnarray}
h(x,p,z)&=(pF_0(x)+p^2D_0(x) )z^2+2(pF_1(x)+p^2D_1(x) )z(1-z)\nonumber \\
& \quad +(pF_2(x)+p^2D_2(x) )(1-z)^2.
\end{eqnarray}
Generalizing the analysis of the two-state model, we introduce the functions
\numparts
\begin{eqnarray}
\label{q0}
q_0&=\beta_+ \left [1-\e^{i \phi}\right ] +\lambda ,\\ 
q_2&= \beta_- \left [1-\e^{-i \phi}\right ] +\lambda,  \\
\lambda&=- \alpha_-\left [1-\e^{i\phi}\right ]-\alpha_+\left [1-\e^{-i\phi}\right ],
\label{q2}
\end{eqnarray}
\endnumparts
and rewrite the action as
\begin{eqnarray}
\fl  S&=\int_{0}^{t}\bigg [p\frac{dx}{dt}-2i \epsilon \phi\frac{dz}{dt} +  q_0(\phi)z^2+q_2(\phi)(1-z)^2  -\lambda(\phi) -h(x,p,z)\bigg ]d\tau .
 \label{3act3}
\end{eqnarray}
Note that after some algebra one finds that $\lambda,q_0,q_2$ are related according to
\numparts 
\begin{eqnarray}
\label{qq0}
 &(q_0-\beta_+)r_0+\alpha_+ r_1=\lambda r_0,\\ & \beta_+r_0-(\alpha_++\alpha_-)r_1+\beta_-r_2=\lambda r_1,\\
&\alpha_- r_1+(q_2-\beta_-) r_2=\lambda r_2.
\label{qq2}
\end{eqnarray}
\endnumparts
This can be rewritten in the matrix form
\begin{equation}
\sum_{m=1}^3 \left \{q_m \delta_{n,m} +Q_{nm}\right \} r_m=\lambda r_n,\quad \mbox{ such that } q_1=0.
\end{equation}

In the limit $\epsilon \rightarrow 0$ the path integral will be dominated by paths that satisfy
\begin{equation}
2 \epsilon \frac{dz}{dt}=\frac{dq_0}{d\phi}z^2+\frac{dq_2}{d\phi}(1-z)^2-\frac{d\lambda}{d\phi}.
\end{equation}
To leading order we can set $dz/dt=0$ in the action (\ref{act2}) and eliminate the independent variable $\phi$ by imposing the condition
\begin{equation}
\label{moom}
\frac{dq_0}{d\phi}z^2+\frac{dq_2}{d\phi}(1-z)^2-\frac{d\lambda}{d\phi}=0.
\end{equation}
In particular,
\begin{equation}
\label{3zoo}
\e^{2i\phi}=\frac{\beta_- (1-z)^2+2\alpha_+z(1-z)}{\beta_+ z^2+2\alpha_-z(1-z)}.
\end{equation}
We thus obtain the reduced path integral (\ref{pired}) with effective action
\begin{eqnarray}
 \widehat{S}&=\int_{0}^{t}\bigg [p\frac{dx}{dt} +q_0 z^2+q_2(1-z)^2  -\lambda-h(x,p,z)\bigg ]d\tau .
 \label{3act3}
\end{eqnarray}
We can now eliminate $z$ by functionally minimizing the action $\widehat{S}$ with respect to $z$, noting that $q_0,q_1$ and $\lambda$ are functions of $z$ via their dependence on $\phi(z)$:
\begin{eqnarray}
\label{3dude}
 \fl 0&=\frac{\delta \widehat{S}}{\delta z(t)}= \frac{dq_0}{d \phi}\frac{d \phi}{dz}z^2+2zq_0+\frac{dq_2}{d\phi}\frac{d \phi}{dz}(1-z)^2-2(1-z)q_2-\frac{d \lambda}{d \phi }\frac{d\phi}{d z} \\
\fl  &\hspace{3cm} -2p[zF_0(x)+(1-2z)F_1(x)-(1-z)F_2(x)]\nonumber \\
\fl &\hspace{3cm} -2p^2[zD_0(x)+(1-2z)D_1(x)-(1-z)D_2(x)].\nonumber
\end{eqnarray}
It then follows from equation (\ref{moom}) that
\begin{equation}
\fl q_0=p(F_0-F_1)+p^2(D_0-D_1),\quad q_2=p(F_2-F_1)+p^2(D_2-D_1).
\end{equation}
Hence,
\begin{eqnarray}
\fl q_0z^2+q_2(1-z)^2-h(z,x,p)&=-[pF_1+p^2D_1][z^2+2z(1-z)+(1-z)^2]\nonumber \\
&=-[pF_1+p^2D_1],
\end{eqnarray}
and the minimized action is given by equation (\ref{act4}).
Finally, defining $\Lambda=\lambda +pF_1+p^2 D_1$ and substituting for $q_0,q_1$ and $\lambda$ into equations (\ref{qq0})--(\ref{qq2}) recovers the eigenvalue equation (\ref{eigR}) for the three-state model with matrix generator (\ref{Q3}).

\section{Discussion}

In this paper we used coherent spin states to derive a new path integral representation of the probability density functional for a stochastic hybrid system evolving according to a piecewise SDE. A Langevin equation was obtained in the semi-classical limit, which extended previous diffusion approximations based on a quasi-steady-state reduction. It was also shown how the path integral reduces to a previous version in the weak noise limit, whose action functional is related to a large deviation principle. In particular, least action paths can be used to determine the most likely paths of escape from a metastable state. 

A natural extension of the current work is to explore what happens when the number of discrete states is large and the transition rates are $n$-dependent. For example, membrane voltage fluctuations in a neuron may be driven by hundreds of stochastic ion channels, with the number $N(t)$ of open ion channels at time $t$ evolving according to a birth-death master equation \cite{Keener11,NBK13}:
\begin{eqnarray}
\label{bdbd}
\fl \frac{dP_n(n)}{dt}&=\sum_{m\geq 0}Q_{nm}P_m(t)\equiv \omega_+(n-1)P_{n-1}(t)
+  \omega_-(n+1)P_{n+1}(t)\nonumber \\
\fl &\quad -[\omega_+(n)+ \omega_-(n)]P_n(t), 
\end{eqnarray}
with transition rates 
\begin{equation}
\label{ion2}
\omega_+(n,x)=(N_T-n)\alpha(x),\quad \omega_-(n,x)=\beta(x).
\end{equation}
Here $N_T$ is the total number of ion channels. Let $x(t)$ denote the membrane voltage at time $t$, which evolves according to the PDMP
\begin{equation}
\label{ion1}
\frac{dx}{dt}=\left [\frac{n}{N_T}f(x)-g(x)\right ],
\end{equation}
One could construct a coherent spin-$S$ decomposition of the discrete master equation along the lines of the two-state and three-state models with $2S+1=N_T$. This could then be used to analyze the resulting stochastic dynamics in the semi-classical limit. However, the expressions become rather cumbersome for large $S$. On the other hand, it is relatively straightforward to calculate the Hamiltonian $\Lambda$ in the path integral representation (\ref{piold}). 
For the given ion channel model, the eigenvalue equation (\ref{eigR}) becomes
\begin{eqnarray}
\fl &   p\left (\frac{n}{N_{T}}f-g\right )R_{n}(x,p)   +\omega_+(n-1)R_{n-1}(x,p)
+  \omega_-(n+1)R_{n+1}(x,p)\nonumber \\
\fl &\quad -[\omega_+(n)+ \omega_-(n)]R_n(x,p),=\Lambda(x,p)R_{n}(x,p)  
\label{Qr}
\end{eqnarray} 
It turns out that the principal eigenvalue can be determined by considering the positive trial solution \cite{Bressloff14}
\begin{equation}
R_n(x,p)=\frac{\Gamma^n(x,p)}{(N_{T}-n)!n!}.
\end{equation}
Substituting into equation (\ref{Qr}) 
yields the following equation relating $\Gamma$ and $\Lambda_0$:
\begin{eqnarray*}
\fl p\left (\frac{n}{N_{T}}f-g\right ) +\frac{n\alpha}{\Gamma}+\Gamma \beta(N_{T}-n)   -n\beta-(N_{T}-n)\alpha =\Lambda .
\end{eqnarray*}
Collecting terms independent of $n$ and terms linear in $n$ yields a pair of equations
for $\Gamma$ and $\Lambda$. After eliminating $\Gamma$, we obtain a quadratic equation for $\Lambda$ of the form
\begin{equation}
\label{ll0}
\Lambda^2+\sigma(x,p)\Lambda-h(x,p)=0,
\end{equation}
with
\begin{eqnarray*}
\fl \sigma(x,p)&=&p[2g(x)-f(x)]+N_{T}(\alpha(x)+\beta(x)),\\ 
\fl h(x,p)&=&p\bigg  [g(x)\bigg (p(f(x)-g(x))-N_{T}(\alpha(x)+\beta(x)\bigg ) +N_{T}\alpha(x)f(x)\bigg ].
\end{eqnarray*}
One of the roots corresponds to the principal eigenvalue. Elsewhere the reulting Hamiltonian system has been used to determine least action paths associated with the noise-induced from a neuron's resting state \cite{Keener11,NBK13}.

Another possible extension of the current study would be to consider diffusion in a randomly switching environment; one mechanism for switching would be stochastically-gated reactions such as adsorption. Mathematically speaking, this process is an infinite-dimensional version of a stochastic hybrid system, in which the piecewise deterministic dynamics is given by a reaction-diffusion equation. One method for analyzing such a system is to discretize space and construct the Chapman-Kolmogorov (CK) equation for the resulting finite-dimensional stochastic hybrid system \cite{Bressloff15a}. One could then use coherent spin-states to construct a path-integral  representation of the lattice system. Retaking the continuum limit would then generate a path integral functional for the hybrid reaction-diffusion model.

\end{document}